\documentstyle[twoside,fleqn,espcrc2]{article}
\def\bea{\begin{eqnarray}}
\def\eea{\end{eqnarray}}

\def\ds{\displaystyle}
\newcommand \dsty \displaystyle

\newcommand{\beq}{\begin{equation}}
\newcommand{\eeq}{\end{equation}}
\newcommand{\Lam}{\bar\Lambda}

\newcommand{\MS}{\overline{MS}}
\newcommand{\qq}{\langle \bar q q \rangle}
\newcommand{\nn}{\nonumber}

\newcommand{\AmS}{{\protect\the\textfont2
  A\kern-.1667em\lower.5ex\hbox{M}\kern-.125emS}}

\title{Dynamical Chiral Symmetry Breaking from Variationally Improved
Perturbative Expansion
}

\author{J.-L. Kneur~\address{Physique Math\'ematique et Th\'eorique,  
UMR CNRS, 
Universit\'e Montpellier II,\\  F-34095 Montpellier Cedex 5 France}}
       
\begin{document}
\setcounter{page}{0}
\begin{flushright}
{\large
PM/97--47 \\
September 1997}
\end{flushright}
\vspace{1cm}

\begin{center}

{\Large\sc {\bf Dynamical Chiral Symmetry Breaking from \\

Variationally Improved
Perturbative Expansion~\footnote{Talk given at the QCD97 Euroconference, 
July 1997, Montpellier France, to appear in the proceedings.}}}

 \vspace{1cm}

{\large J.-L. Kneur}

\vspace{1cm}

{\large Physique Math\'ematique et Th\'eorique, UMR--CNRS, \\
Universit\'e Montpellier II, F--34095 Montpellier Cedex 5, France.}

\end{center}

\vspace*{2cm}

\noindent 
{\large
We review how a specific  
resummation of the so-called 
``delta-expansion", applied to the QCD Lagrangian, 
transforms the ordinary perturbative expansion 
in $\alpha_s$ into an expansion
in an arbitrary  mass parameter, around  
the basic scale $\bar\Lambda$. 
When applied to the pole mass, the resulting expression 
may be interpreted as a dynamical mass ansatz, to be optimized
with respect to the new expansion (mass) parameter. The 
construction is generalized to obtain
estimates of the order parameters 
of the $SU(n_f)_L \times SU(n_f)_R$ ($n_f =2,3$) symmetry breakdown.}

\newpage
\setcounter{footnote}{0}
\begin{abstract}
 We review how a specific  
resummation of the so-called 
``delta-expansion", applied to the QCD Lagrangian, 
transforms the ordinary perturbative expansion 
in $\alpha_s$ into an expansion
in an arbitrary  mass parameter, around  
the basic scale $\bar\Lambda$. 
When applied to the pole mass, the resulting expression 
may be interpreted as a dynamical mass ansatz, to be optimized
with respect to the new expansion (mass) parameter. The 
construction is generalized to obtain
estimates of the order parameters 
of the $SU(n_f)_L \times SU(n_f)_R$ ($n_f =2,3$) symmetry breakdown.
\end{abstract}

\maketitle

\section{Introduction}

There is a common lore that a ``first principle" determination of
the order parameters 
characterizing the (chiral) dynamical symmetry
breaking (DSB), such as the $\qq$ condensate typically, 
is definitively out of 
the reach of the basic QCD perturbation theory. This is  
largely justified, traditionally, by the fact that DSB is an
essentially non-perturbative mechanism. However, it may depend 
on what exactly one
means by perturbation theory. 
For instance, 
since the pioneering work of Nambu and Jona-Lasinio
(NJL)~\cite{NJL}, it has been understood how 
it is possible to
resum a relevant class of graphs to obtain 
the qualitative (and some quantitative
as well) properties of DSB explicitly,
at least in specific approximations and/or models. 
Also, independently of the NJL idea, the modern 
Chiral Perturbation Theory (ChPT)~\cite{GaLeut} 
gives a consistent effective description of data at low energies  
where the QCD perturbative series is not applicable.
Indeed, definite progress have been made to relate ChPT with
generalized NJL models~\cite{ENJL}, although 
a precise connection between the (numerous) ChPT parameters, and 
the  basic QCD coupling and quark mass 
parameters is far
from being resolved at present. 
With a more formal (but related) motivation,  it has been also 
explored since long ago how 
definite  non-perturbative informations 
may be inferred from
appropriately modified perturbation 
series~\cite{delta}, at least in simplified or exactly solvable 
models. In particular, 
the convergence of ordinary perturbation
can be systematically {\em improved} 
by a variational--like procedure, in which
the separation of the
action into ``free" and ``interaction" parts
is made to depend on a set of auxiliary parameters, 
to be fixed by some optimization procedure~\footnote{
In $D =1$ field theories, this optimized perturbation
theory (``delta-expansion") 
gives a rigorously convergent~\cite{converg}
series of approximations, even in strong coupling cases.}.

As a partial attempt to merge some of these ideas, 
we have re-examined~\cite{qcd1,qcd2} with a new approach 
the above mentioned old problem of generating from
the basic QCD Lagrangian 
non-trivial values for the 
quark condensate, pion decay constant, or  
dynamical quark
masses.
The basic point is to transform the 
ordinary perturbative expansion, in $\alpha_s$, into an expansion
in an {\em arbitrary}  mass parameter, around  
a non-trivial (fixed-point) 
solution of the renormalization group evolution,
proportional to the basic scale $\bar\Lambda$. 
In some sense it may be viewed as a systematic, order by order, 
improvement of the
original NJL construction,  but with  
a consistent treatment of the renormalization (and directly applied
to the QCD quark-gluon interactions). 
\section{A crude dynamical mass ansatz}
As a  crude first illustration of the mechanism, consider
the renormalization group (RG) evolution of the running mass,  
\beq
m(\mu^{'}) = m(\mu )\;\; {exp\{ -\int^{g(\mu^{'})}_{g(\mu )}
dg {\gamma_m(g) \over {\beta(g)}} \} }\;,
\label{runmass}
\eeq
where $\beta(g)$, $\gamma_m(g)$ drive the running of the coupling 
$g(\mu)$ and mass $m(\mu)$, respectively. Solving (\ref{runmass}) for the 
``fixed point" boundary condition:
$ M \equiv m(M)$,
gives (to first RG order)
\beq
M_1 = {m(\mu) \over{[1 +2b_0 g^2(\mu) \ln({M_1 \over{\mu}})
]^{\gamma_0 \over{2b_0}}}}\;, 
\label{MRG1}
\eeq
where $b_0$, $\gamma_0$ are the one-loop RG-coefficients (normalization 
is such that $\beta(g) = -b_0 g^3 -b_1 g^5 -\cdots$,
$\gamma_m(g) = \gamma_0 g^2 +\gamma_1 g^4
+\cdots$). \\

Although expression (\ref{MRG1}) is initially related 
to the ``current" mass $m(\mu)$ via (\ref{runmass}), it 
has the trademarks
of a {\em pole} mass, thanks to the boundary 
condition $ M_1 \equiv m(M_1)$~\footnote{In particular~\cite{qcd2}, 
(\ref{MRG1}) 
is {\em scale} invariant (in contrast with $m(\mu)$)
and gauge-invariant, 
as the pole mass should be.}.
This coincidence between the 
 pole mass $M$ and the current mass
$m(\mu \equiv M)$, is, however, only an artifact 
of our crude approximation, neglecting at the moment the non-logarithmic
perturbative corrections~\cite{Broadhurst}. 
Now, the most important
property of expression (\ref{MRG1})
is that it is 
{\em non-zero} in the chiral limit, $m(\mu) \to 0$.  
Indeed, (\ref{MRG1}) identically reads 
\beq
M_1 (\ln (M_1/\Lam) )^{\frac{\gamma_0}{2b_0}} = \hat m
\label{M1rewritten}
\eeq
where for convenience we introduced 
the RG invariant scale $\Lam =
\bar\mu \, e^{-{1\over{2b_0 \bar g^2}}}$
(at first RG order), and the {\em scale-invariant} mass
$\hat m \equiv  m(\bar\mu) (2b_0 g^2(\bar\mu))^{-\frac{\gamma_0}{2b_0}} $. 
(\ref{M1rewritten}) may then be seen as a function
$\hat m(M_1)$, and requiring its inverse, $M_1(\hat m)$, to be defined
on the whole physical domain $ 0 < \hat m < \infty$, and to match the 
ordinary perturbative asymptotic behavior for $ \hat m \to \infty$,
implies~\footnote{ 
Another, a priori possible solution of
(\ref{M1rewritten}),  
 $M_1 \to 0$ for $\hat m \to 0$, is rejected  because it 
is only defined for $0 \le \vert \hat m \vert \le 
(\gamma_0/2b_0)^{\gamma_0/2b_0}
e^{-\gamma_0/2b_0} \Lam < \Lam$, and is therefore not compatible with the
asymptotic perturbative behavior of (\ref{MRG1}) for $m(\mu) \gg
\Lam $~\cite{qcd2}.} $M_1 (\hat m \to 0) \to \Lam $. 
It is of course desirable to go beyond the one-loop RG approximation,
and to take into account as well the non-logarithmic 
corrections, necessary to make contact with the usual perturbative
pole mass~\cite{Broadhurst}. 
Our aim is    
 to obtain a variational ``mass gap" where the 
 non-trivial chiral limit property of (\ref{MRG1}) 
is preserved, while
at the same time providing us with a systematically (order by order)
 improvable ansatz, thanks to a particular reorganization of the
basic perturbative expansion, as will be explained in the
 next section.

\section{Resumming the delta-expansion} 

In the present context, a simplest form of the so-called 
delta-expansion~\cite{delta} may
 be defined  
by formally substituting everywhere in the bare QCD Lagrangian:
\beq m_0 \to m_0\; (1-x); ~~~~g_0 \to g_0\; (x)^{1/2}\;.
\label{substitution}
\eeq
 The
 parameter $x$ in 
(\ref{substitution}) just 
interpolates between the free Lagrangian, for
$x=0$, and the interacting but {\em massless} Lagrangian, for $x=1$.    
In the simplest field-theoretical applications, one would then  use
(\ref{substitution}) to expand any perturbative expression of 
($m_0$, $g^2_0$)
to a given order $x^q$, and try to 
apply some optimization
prescription with respect to the (arbitrary) mass,
$m_0$. Accordingly, the somewhat empirical 
but most often successful idea~\cite{delta} is that the 
{\em least sensitive} region with respect to $m_0$ (entering at any
fixed order $q$)  
should give the best approximation to the exact result, which is 
{\em independent} of
$m_0$. But, in many non-trivial field theories, and in particular  
in the present QCD case, before anything the whole procedure 
should be made consistent with renormalization.  As it turns out, the only
way to get
a finite and non-zero result (e.g, $M(m \to 0) \neq 0$) is 
to resum 
the $x$-series, using  
an appropriately constructed 
contour integral transform~\cite{gn2}. At first RG order,   
this essentially gives a mass as an integral over
expression (\ref{MRG1}) (with substitution (\ref{substitution})
understood). Beyond the one-loop approximation,   
our final mass ansatz reads~\footnote{$v$ in (\ref{contour7}) 
is related to the original
expansion parameter $x$ as $x = 1-v/q$, $q$
being the order of the $x$-expansion.}: 
\bea
{ M^P_2 (m^{''})\over \Lam}
 = {2^{-C} m''\over{2 i \pi}} \oint dv {e^{\;v} 
\over{F^A(v) [C + F(v)]^B}}  \nn \\
\cdot {\left(1 +{{\cal M}_{1}\over{F(v)}}
+{{\cal M}_{2}\over{F^2(v)}}+\cdots \right)},
\label{contour7}
\eea
where the contour is around the $] -\infty, 0] $ axis; 
\beq
F(v) \equiv \ln [m''v] -A \; \ln F -(B-C)\; \ln [C +F],
\label{Fdef}
\eeq
with $A =\gamma_1/(2 b_1)$, $B =\gamma_0/(2 b_0)-\gamma_1/(2 b_1)$,
$C = b_1/(2b^2_0)$; 
$\Lam$ is the (RG-invariant) scale 
at two-loop order;
and finally
\beq
m''\equiv  \ds{\left(\frac{m(\bar\mu)}{ \Lam}\right) \;
2^{C}\;[2b_0 \bar g^2]^{-\frac{\gamma_0}{2b_0}}
\;\left[1+\frac{b_1}{b_0}\bar g^2\right]^B}
\; 
\label{msec2def}
\eeq
is the scale-invariant, arbitrary (dimensionless) ``mass" parameter.
By construction, $F(1)$ in the integrand of (\ref{contour7}) 
resums the leading and next-to-leading logarithmic
dependence in $m(\bar\mu)$ to all orders~\cite{qcd2}.
The non-logarithmic perturbative coefficients, 
${\cal M}_{1} \equiv (2/3)~(\gamma_0/2b_0)$ and ${\cal M}_2$,  
connect~\cite{Broadhurst} the pole mass with 
the running mass $m(M)$. \\

Note that it is implicitly always possible to choose a 
renormalization scheme (RS) such that $b_i = \gamma_i =0$ for
$i \geq 2$, since $b_i$, $\gamma_i$ are then RS--dependent. In that sense,  
eq.~(\ref{contour7}) resums
the full RG dependence in $\ln (m'' v)$. 
In contrast, the purely perturbative (non-logarithmic)
information, contained in ${\cal M}_{1}$, ${\cal M}_{2}$, is
limited by present knowledge to two-loop order. 
This is where the variational principle
and optimization play their role, whereby we hope to obtain a sensible
approximation to the true dynamical mass.
Observe first 
that, were we in a simplified theory where 
${\cal M}_{1} = {\cal M}_{2} = \cdots = 0$, 
(\ref{contour7}) would have a very simple
behaviour near its optimum (located at $m'' \to 0$), 
giving a simple pole 
with residue $M_2 = (2C)^{-C}\;\Lam $.
Now, in the more realistic cases, ${\cal M}_1$, ${\cal M}_2$,... 
cannot be neglected, 
but we can obtain   
a series of approximants to the dynamical mass, 
by expanding (\ref{contour7}) in successive powers of $m'' v$, 
 using the standard relation
\beq
\label{hankel}
\frac{1}{2i \pi} \oint dv\: e^v \: v^\alpha  = 
\frac{1}{\Gamma[-\alpha]}\; ,
\eeq
and then looking for optima $M^P_2(m''_{opt})$, $m''_{opt} \neq 0$. \\
The previous construction is quite general and therefore 
directly applicable to any (asymptotically free) model, 
taking obviously its appropriate values of the RG coefficients,
$b_i$, $\gamma_i$.
The ansatz (\ref{contour7}) was 
confronted~\cite{gn2} to the exactly known mass
gap~\cite{FNW} 
for the ${\cal O}(N)$ Gross-Neveu (GN) model, for {\em arbitrary}
$N$. 
The results
of different optimization prescriptions gave
estimates with errors of ${\cal O}$(5\%) or less, depending
on $N$ values~\cite{gn2}. 
It is important to note 
that expression (\ref{contour7}), for arbitrary $N$ in the
GN model, uses exactly the {\em same} 
amount of (perturbative plus RG) information than the one
at disposal at present for a QCD quark mass: namely,
the {\em exact} two-loop RG-resummed 
plus perturbative ${\cal M}_1$, ${\cal M}_2$ 
dependence.
Since our construction essentially relies on RG-properties (and 
analytic continuation), going from 2 to 4 dimensions 
is not expected to cause major changes, at least naively. 
\section{Hidden singularities of the mass ansatz}
One complication, actually, {\em does} occur: 
as a more careful examination of 
relation (\ref{Fdef}) indicates, there are  
branch cuts 
in the $v$ plane, with ${\rm Re}[v_{cut}] > 0$
for the relevant case of $n_f =$ 2 or 3 in QCD.
These make the expansion undefined when
approaching the origin, $v=0$, and simply indicate 
the non single--valuedness of (\ref{contour7})
below those branch points.    
The origin of those singularities has some 
similarity with the renormalon ones~\cite{renormalons}, 
as they also appear  
when extrapolating a RG--resummed expression 
down to an infrared scale $m'' \simeq 0$.
However, a main difference with renormalons is that
in our construction
it is possible~\cite{qcd2} to move those extra cuts to 
a safe location, ${\rm Re}[v^{'}_{cut}] \leq 0$, observing that
the actual position of those cuts
depends, at second order, on the RS, via
$\gamma_1$.  Performing 
thus a second-order 
perturbative RS change in $m(\mu)$, $g(\mu)$, 
which changes $\gamma_1(\MS)$ to a (singularity-safe) $\gamma^{'}_1$, 
it is then sensible, in the present context, 
to invoke a variant of the ``principle of
minimal sensitivity" (PMS)~\cite{delta}, requiring
a flat optimum (plateau) of (\ref{contour7})
with respect to the
{\em remnant} RS arbitrariness~\cite{qcd2}. \\
One may perhaps legitimately wonder why the ordinary
renormalon singularities of the pole mass~\cite{BeBr94}
 do not seem to appear
in our construction. In fact, the usual renormalon
singularities always appear as a result of crossing  
the Landau pole~\footnote{Actually, this is an oversimplified
picture, valid at one-loop
RG level only~\cite{EdRPe96}. 
However, higher order properties of renormalons  
do not affect, qualitatively, our argument.}, 
which simply reflects an ambiguity from perturbation theory, calling for
non-perturbative corrections which are typically in the form of
power corrections~\cite{renormalons}. 
In contrast, (\ref{contour7}) is such that the Landau pole
(corresponding to $F =0$ in our language) is {\em not} crossed, but only
smoothly reached from above, ${\rm Re} F >0$. 
(Moreover, due to the recurrent dependence
in $F$, (\ref{Fdef}), implying that $F(v) \simeq m^{''} v$ for
$m^{''} v \to 0$, the poles of (\ref{contour7}) at $F=0$ ($v =0$) 
entirely come
from the purely 
perturbative part, i.e. due to ${\cal M}_1, {\cal M}_2 \neq 0$).
Note that, on more
phenomenological grounds, there is no strong contradiction with the usual
consequences of the presence of renormalons: while the latter indicate,
in the pole mass case, 
an ambiguity of ${\cal O}(\Lam)$~\cite{BeBr94},
our construction necessarily exhibits an {\em arbitrary} renormalization
scheme (RS) dependence, via the above mentioned 
$\gamma_1$ coefficient, calling for optimization. 
Practically we have obtained:
\beq
M^2_{opt}(m''_{opt} \to 0) \simeq 2.97\;\Lam(2)\;
\label{Mnum}
\eeq
for $n_f=2$, and a similar result for $n_f =3$.
\section{Order parameters: $F_\pi$ and $\qq$}
The previous dynamical quark mass, although it has some meaning 
as regards DSB in QCD, hardly has a direct physical
interpretation, e.g. 
as a pole of the S-matrix, due to the confinement.
In other words, it is not a properly defined order parameter. 
It is however possible to apply the same construction as the one
leading to 
(\ref{contour7}), 
to obtain  
a determination of the ratios $F_\pi/\Lam$ and $\qq(\mu)/\Lam^3$.
The latter gauge-invariant quantities are 
unambiguous order parameters, i.e. $F_\pi \neq 0$ {\em or} $\qq \neq 0$ 
{\em imply} DSB. 
The appropriate generalization
of (\ref{contour7}) for $F_\pi$ is~\cite{qcd2}
\bea
& \ds{{F^2_\pi \over{\Lam^2}} = (2b_0)\;
{2^{-2 C} (m'')^2\over{2 i \pi}} \oint {dv\over v}\; v^2 {e^{\: v}}}
\; \nn \\
& \ds{ \cdot \;\frac{1}{F^{\;2 A-1} [C + F]^{\;2 B}} 
 \; \delta_{\pi }
 \left(1 +{\alpha_{\pi}\over{F}}+{\beta_{\pi}
\over{F^2}}
\right) }
\label{Fpiansatz}
\eea
in terms of the same 
$F(v)$ defined in eq.~(\ref{Fdef}) (therefore leading to the same
extra cut locations as in the mass case), and where
$\delta_\pi$, $\alpha_\pi$ and $\beta_\pi$
are fixed by matching the perturbative $\MS$
expansion, known to 3-loop order~\cite{Avdeev}. 
A numerical optimization with respect to the RS-dependence, in a way
similar to the mass case, gives e.g for $n_f =2$: 
\beq
F_{\pi ,opt}(m''_{opt} \to 0)
 \simeq 0.55\;\Lam(2)\;.
\label{Fpinum}
\eeq
Concerning $\qq$, an ansatz similar 
to (\ref{Fpiansatz}) can be derived (with 
coefficients $\delta$, $\alpha$, $\beta$ 
specific to $\qq$ and appropriate changes in the 
$m''$, $F$ and $v$ powers),
but for the RG-invariant combination $m \qq$,
due to the fact that our construction  
only apply to RG-invariant quantities.
 To extract an estimate of 
the (scale-dependent) condensate $\qq(\mu)$ is only  
possible by introducing 
an {\em explicit} symmetry-breaking
quark mass $m_{exp}$ (i.e.  
$m_{exp} \neq m $), 
and expanding the $m\qq$ ansatz to first order in $m_{exp}$. 
This gives
for $n_f =2$~\cite{qcd2}: 
\beq
\qq^{1/3} (\bar\mu = 1\;\mbox{GeV}) \simeq 0.52 \; \Lam(2)\;.
\label{qqnum}
\eeq 
Confronting (\ref{Mnum}), (\ref{Fpinum}) and (\ref{qqnum})
gives a fairly small value of the quark condensate~\footnote{The smallness
of $\qq$ is however essentially correlated with the smallness of the
$F_\pi/\Lam$ ratio estimate in our framework, eq.~(\ref{Fpinum}).} 
(and a fairly
high value of the dynamical mass), as compared to other non-perturbative
determinations~\cite{sumrules}.
Although small values of $\qq$ are 
not experimentally excluded
at present~\cite{Stern}, it is also clear that our relatively crude 
approximation deserves more refinements 
for more realistic QCD predictions.
\section{Conclusion and discussion}
The variationally improved expansion in arbitrary $m''$, first developed
in the GN model~\cite{gn2}, has been formally
extended to the QCD case.
It gives non-trivial relationships between
$\Lam$ and the dynamical masses and order parameters, $F_\pi$
and $\qq$.

To make progress, what is certainly restrictive is the 
relatively poor knowledge
of the purely perturbative part of the expansion (only known to 
two-loop order in most
realistic field theories).  Accordingly, our final numerical results
crucially depend on the optimization~\footnote{ 
For instance, results for (\ref{contour7}), (\ref{Fpiansatz}) 
are substantially different~\cite{qcd2} 
in the unoptimized $\MS$ scheme.}. Apart from a few
models where the series is known to large orders
(as in the
anharmonic oscillator~\cite{delta,bgn}, 
or in the GN model for $N \to \infty$),
we can hardly compare successive orders of this
expansion to estimate, even qualitatively, the
{\em intrinsic} error of such a method.
Invoking the PMS principle~\cite{delta}, although physically 
motivated, may  
artificially 
force the series to converge, with no guarantees that it is 
toward the right result. 
\section*{Questions}
\noindent {\bf E. de Rafael}: Can one simply see, just at first RG order
approximation already, what type of graphs 
are resummed by this mass ansatz? \\
{\bf J.L.K}: It contains the ``bubble" chain (the one-loop
insertions in the gluon line) but, in addition, 
there is an iteration of those dressed gluon lines (the so-called
Ladder graphs). \\
{\bf S. Narison}: Your calculation is essentially
perturbative. Can you include in it the truly non perturbative contributions,
like condensates as they appear in the 
operator product expansion typically? \\
{\bf J.L.K}: the aim here is to try to 
estimate these NP quantities from the basic QCD interactions only, 
with of course this peculiar resummation.  
To include OPE-like condensates from the start would
be a kind of double-counting.
\end{document}